\documentclass[preprint,pre,aps,showpacs]{revtex4}
\usepackage{graphicx}
\begin{document}

\title{Improvements to model of projectile fragmentation}

\author{S. Mallik, G. Chaudhuri}
\affiliation{Variable Energy Cyclotron Centre, 1/AF Bidhannagar,
Kolkata 700064, India}
\author{S. Das Gupta}
\affiliation{Physics Department, McGill University,
Montr{\'e}al, Canada H3A 2T8}

\date{\today}

\begin{abstract}
In a recent paper \cite{Mallik} we proposed a model for calculating cross-sections of
various reaction products which arise from disintegration of projectile like
fragment resulting from heavy ion collisions at intermediate or higher energy.
The model has three parts: (1) abrasion, (2) disintegration of the hot abraded
projectile like fragment (PLF) into nucleons and primary composites using
a model of equilibrium statistical mechanics
and (3) possible evaporation of hot primary composites.  It was assumed that
the PLF resulting from abrasion has one temperature $T$.
Data suggested that while just one value of $T$ seemed adequate for most
cross-sections calculations,
it failed when dealing with very peripheral collisions.  We have now introduced
a variable $T=T(b)$ where $b$ is the impact parameter of the collision.  We
argue there are data which not only show that $T$ must be a function of $b$
but, in addition, also point to an approximate value of $T$ for a given $b$.
We propose a very simple formula: $T(b)=D_0+D_1(A_s(b)/A_0)$
where $A_s(b)$ is the mass of the abraded PLF
and $A_0$ is the mass of the projectile; $D_0$ and
$D_1$ are constants.  Using this model we compute cross-sections
for several collisions and compare with data.

\end{abstract}


\maketitle
\section{Introduction}

In a recent paper \cite{Mallik} we proposed a model of projectile mutifragmentation
which was applied to collisions of Ni on Be and Ta at 140 MeV/nucleon and Xe
on Al at 790 MeV/nucleon.  The model gave reasonable answers for most of the
cross-sections studied.  The model requires
integration over impact parameter.  For a given impact parameter,
the part of the projectile that does not directly overlap with the target
is sheared off and defines the projectile like fragment(PLF).  This is abrasion
and appealing to the high enrgy of the beam, is calculated using straight
line geometry.  The PLF has $N_s$ neutrons,$Z_s$ protons and $A_s(=N_s+Z_s)$
nucleons (the corresponding quantities for the full projectile are labelled
$N_0,Z_0$ and $A_0$).  The abraded system $N_s,Z_s$ has a temperture.
In the second stage this hot PLF expands to one-third of the normal nuclear
density.  Assuming statistical equilibrium the break up of the PLF
at a temperature $T$ is now
calculated using the canonical thermodynamic model(CTM). The composites that
result from this break up have the same temperature $T$ and can evolve further
by sequential decay(evaporation).  This is computed.  Cross-sections can
now be
compared with experiment.  The agreements were reasonable except for
very peripheral collisions and it was conjectured in \cite{Mallik} that the main reason
for this discrepancy was due to the assumption of constant $T$ over all
impact parameters.

Full details are provided in \cite{Mallik}.  Our aim here is to improve the model by
incorporating an impact parameter dependence of $T=T(b)$.  While we were
led to this by computing the cross-sections of very large PLF's (which can only
result from very peripheral collisions), the effect of temperature dependence
is accentuated in other experiments.  In fact these experiments can be used,
with some aid from reasonable models, to extract ``experimental'' values
for temperature $T$ at each $b$.  We spend considerable time studying this
although our primary aim was and is the computation of cross-sections from
a theoretical model.

\section{Basics of the Model}
Consider the abrasion stage.  The projectile hits the target.  Use straight
line geometry.  We can then calculate the volume of the projectile that
goes into the participant region (eqs.A.4.4 and A.4.5 of ref \cite{Dasgupta1}).  What
remains in the PLF is $V$.  This is a function of $b$.  If the original
volume of the projectile is $V_0$, the original number of neutrons is $N_0$
and the original number of protons is $Z_0$, then the average of neutrons
in the PLF is $<N_s(b)>=[V(b)/V_0]N_0$ and the average number of protons
is $<Z_s(b)>=[V(b)/V_0]Z_0$;
$<N_s(b)>$ (and similarly $<Z_s(b)>$) is usually a non-integer.  Since in
any event only an integral number of neutrons (and protons) can appear
in a PLF we need a prescription to get integral numbers.
Let the two nearest integers to $<N_s(b)>$ be
$N_s^{min}(b)$ and $N_s^{max}(b)=N_s^{min}(b)+1$.  We assume that
$P_{N_s}(b)$=the probability that the abraded system has $N_s$ neutrons
is zero unless $N_s(b)$is either $N_s^{min}(b)$ or $N_s^{max}(b)$.
Let $<N_s(b)>=N_s^{min}(b)+\alpha$ where $\alpha$ is less than 1.  Then
$P(N_s^{max}(b))=\alpha$ and $P(N_s^{min}(b)=1-\alpha$.
Similar condions apply to $P_{Z_s}(b)$.  The probability that a PLF with
$N_s$ neutrons and $Z_s$ protons materializes from a collision at impact
parameter $b$ is given by $P_{N_s,Z_s}(b)=P_{N_s}(b)P_{Z_s}(b)$.  Once this
PLF is formed it will expand and break up into composites at a temperature
$T$.  We use CTM to obtain these.  All the relevant details of CTM can be
found in \cite{Mallik} and \cite{Das}.  We will not repeat these here.
There can be very light fragments, intermediate mass fragments
(defined more precisely in the next section) and heavier fragments.
As the fragments are at temperature $T$ it is possible some of these will
sequentially decay thereby changing the final population which is measured
experimentally.  Details of evaporation can be found in \cite{Mallik} and \cite{Chaudhuri1}.
\section{Arguments for $b$-dependence of Temperature}
Experimental data on $M_{IMF}$ as a function of $Z_{bound}$ (see Fig.1 in \cite{Sfienti})
probably provide the strongest arguments for needing an impact parameter
dependence of the temperature. Here $M_{IMF}$ is the average multiplicity
of intermediate mass fragments (in this work those with $z$ between 3 and
20) and $Z_{bound}$=sum of all charges coming from PLF minus particles with
$z$=1.  For ease of arguments we will neglect, in this section, the
difference between $Z_{bound}$ and $Z_s$, the total charge of all
particles which originate from the PLF.  A large value of $Z_{bound}$
(close to $Z_0$ of the projectile) signifies that the PLF is large and
the collision is peripheral (large $b$)
whereas a relatively smaller  value of $Z_{bound}$
will imply more central collision (small $b$).
For equal mass collision $Z_{bound}$
goes from zero to $Z_0$, the total charge of the projectile.

The following gross features of heavy ion collisions at intermediate
energy are known.  If the excitation energy (or the temperature) of
the dissociating system is low then one large fragment and a small
number of very light
fragments emerge.  The average multiplicity of IMF is very small.
As the temperature increases,
very light as well as intermediate mass fragments appear at the expense
of the heavy fragment. The multiplicity $M_{IMF}$ will grow as a function
of temperature, will reach a peak and then begin to go down as, at a high
temperature, only light particles are dominant.  For evidence and discussion
of this see \cite{Tsang}.
For projectile fragmentation
we are in the domain where $M_{IMF}$ rises with temperature.  Now at constant
temperature, let us consider what must happen if the dissociating system
grows bigger.  We expect $M_{IMF}$ will increase with the size of the
dissociating system, that is, with $Z_{bound}$. Experimental data are
quite different: $M_{IMF}$ initially increases, reaches a maximum at a
particular value of $Z_{bound}$ and then goes down.

In Fig.1 we show two graphs for $M_{IMF}$, one in which the temperature
is kept fixed (at 6.73 MeV) and another in which $T$ decreases linearly
from 7.5 MeV (at $b$=0)to 3 MeV at $b_{max}$.  The calculation is qualitative.
The case considered is $^{124}$Sn on $^{119}$Sn.  CTM is used to calculate
$M_{IMF}$ but evaporation is not included.  Simlarly $Z_{bound}$ is
$Z_s$ (no correction for $z=1$ particles).  Fuller calculations will be
shown later but the principal effects are all in the graphs.  Keeping the
temparature fixed makes $M_{IMF}$ go up all the way till $Z_{bound}$=$Z_0$
is reached.  One needs the temperature to go down to bring down the value
of $M_{IMF}$ as seen in experiment.

\section{Use a model to extract $b$-dependence of Temperature}
In our model we can use an iterative technique to deduce a temperature
from experimental data of $M_{IMF}$ $\it{vs}$ $Z_{bound}$.  Pick a $b$;
abrasion gives a $<Z_s>$.  Guess a temperature $T$. A full calculation
with CTM and evaporation is now done
to get a $Z_{bound}$ and $M_{IMF}$.  This $Z_{bound}$ will be
close to $<Z_s>$.  If the guessed value of temperature is too low
then the calculated value of $M_{IMF}$ will be too little for this value
of $Z_{bound}$ when confronted with data.  In the next iteration the
temperature will be raised.  If on the other hand, for the guess
value of $T$, the calculated $M_{IMF}$ is too high, in the next iteration
the temperature will be lowered.
Of course when we change $T$, calculated $Z_{bound}$
will also shift but this change is smaller and with a small number
of iterations one can approximately reproduce an experimental pair
$Z_{bound},M_{IMF}$.
\begin{table}[h]
\begin{center}
\begin{tabular}{|c|c|c|c|c|c|}
\hline
\multicolumn{2}{|l|}{Experimental} & \multicolumn{4}{c|}{Theoretical} \\
%\multicolumn{4} {|1|} {expt} & theory\\
\hline
$Z_{bound}$ & $M_{IMF}$ & $Z_{bound}$ & $M_{IMF}$ & b & Required T\\
 & & & & (fm) & (MeV)\\
\hline
11.0 & 1.421 & 11.080 & 1.424 & 2.912 & 6.398\\
\hline
15.0 & 1.825 & 15.094 & 1.818 & 3.625 & 6.108\\
\hline
20.0 & 2.145 & 19.984 & 2.131 & 4.4574 & 5.840\\
\hline
25.0 & 2.010 & 25.024 & 2.019 & 5.289 & 5.520\\
\hline
30.0 & 1.505 & 29.854 & 1.545 & 6.122 & 5.250\\
\hline
35.0 & 0.920 & 34.985 & 0.928 & 7.072 & 4.970\\
\hline
40.0 & 0.415 & 39.639 & 0.424 & 8.023 & 4.650\\
\hline
45.0 & 0.193 & 44.763 & 0.196 & 9.331 & 4.350\\
\hline
47.0 & 0.156 & 46.512 & 0.154 & 9.925 & 4.260\\
\hline
49.0 & 0.135 & 48.425 & 0.130 & 10.876 & 4.190\\
\hline
\end{tabular}
\end{center}
\caption{ Best fit and experimental values for $^{124}$Sn on $^{119}$Sn. The first two columns are data from experiment.  The next two columns
are the values of $Z_{bound}$ and $M_{IMF}$ we get from our iterative
procedure.  These values are taken to be close enough to the experimental pair.
These are obtained for a value of $b$ (fifth column) and a temperature $T$
(sixth column).}
\end{table}\\

For the case of $^{124}$Sn on $^{119}$Sn we provide Table I which demonstrates
this.  The first two columns are data from experiment.  The next two columns
are the values of $Z_{bound}$ and $M_{IMF}$ we get from our iterative
procedure.  These values are taken to be close enough to the experimental pair.
These are obtained for a value of $b$ (sixth column) and a temperature $T$
(fifth column).  Table II provides similar compilation for $^{107}$Sn
on $^{119}$Sn.
\begin{table}[h]
\begin{center}
\begin{tabular}{|c|c|c|c|c|c|}
\hline
\multicolumn{2}{|l|}{Experimental} & \multicolumn{4}{c|}{Theoretical} \\
%\multicolumn{4} {|1|} {expt} & theory\\
\hline
$Z_{bound}$ & $M_{IMF}$ & $Z_{bound}$ & $M_{IMF}$ & b & Required T\\
 & & & & (fm) & (MeV)\\
\hline
15.0 & 1.690 & 14.816 & 1.583 & 3.886 & 6.200\\
\hline
20.0 & 1.923 & 19.865 & 1.906 & 4.698 & 5.740\\
\hline
21.0 & 1.984 & 21.207 & 1.976 & 4.930 & 5.705\\
\hline
25.0 & 1.749 & 24.913 & 1.758 & 5.510 & 5.320\\
\hline
30.0 & 1.079 & 30.356 & 1.075 & 6.438 & 4.900\\
\hline
35.0 & 0.581 & 35.252 & 0.602 & 7.366 & 4.600\\
\hline
40.0 & 0.223 & 40.123 & 0.225 & 8.410 & 4.210\\
\hline
45.0 & 0.201 & 44.676 & 0.199 & 9.802 & 4.100\\
\hline
47.0 & 0.201 & 47024 & 0.159 & 10.876 & 4.000\\
\hline
\end{tabular}
\end{center}
\caption{ Same as Table 1, except that here the projectile is $^{107}$Sn instead of $^{124}$Sn.}
\end{table}\\
Having deduced once for all such ``experimental data'' of
$T$ $\it{vs}$ $b$, one can try simple parametrisation
like $T(b)=C_0+C_1*b+C_2*b^2...$.
and see how well they fit the data.  We show this for the two cases
in Fig.2.

In Fig.3 using such parametrised versions of $T$ we compute
$M_{IMF}$ $\it {vs}$ $Z_{bound}$ and compare with experimental data.  Except
for fluctuations in the values of $M_{IMF}$ for very low values of $Z_{bound}$
the fits are very good.  We will return to the cases of fluctuations in a
later section.

\section{Temperatures extracted from isotope populations}
In the preceding sections we have extracted temperatures $T$ (combining data
and model) at values of $b$ (equivalently at values of $Z_{bound}$).  This
is a new method for extracting temperature.  A more standard way of extracting
temperatures is the Albergo formula \cite{Albergo} which has been widely used in the
past (for a review see, for example, \cite{Dasgupta2,Pochodzalla}).  In [\cite{Ogul}, Figs.24 and 25]
temperatures at selected values of $Z_{bound}/Z_0$ were extracted from
populations in [$^{3,4}$He,$^{6,7}$Li] and [$^{7,9}$Be,$^{6,8}$Li] using
Albergo formula.
These temperatures are compared in Fig.4 with a typical temperature profile
deduced here.  It is gratifying to see that such different methods of
extraction still give reasonable agreement.

\section{Fluctuations in $M_{IMF}$ for small $Z_{bound}$}
For small values of $Z_{bound}$ the measured $M_{IMF}$ shows considerable
fluctuations as we go from one value of $Z_{bound}$ to another (see Fig.3).
Our model does not reproduce these although general shapes are correct.
Statistical models are not expected to show such fluctuations but let us
get into some details which (a) give a clue how such fluctuations may
arise and (b) why our model misses them.  The reader who is not interested
in such details can skip the rest of this section without loss of continuity.

For definiteness, consider the
case of $^{124}$Sn on $^{119}$Sn.  By the definition of $IMF,(z>2)$, $M_{IMF}$
is 0 when $Z_{bound}$ is 2.  Consider now $Z_{bound}=3$.  The most direct
way one can have this is if the PLF has $Z_s$=3.  Taking very simplistic
point of view, suppose, this also has $N_s=3$, that is, the PLF is $^6$Li.
This
is stable and we immediately get $M_{IMF}=1$.  This is indeed the experimental
value.  The case of $Z_{bound}$=4 may arise if the PLF is $Z_s=4,N_s=4$, i.e.,
if the PLF is $^8$Be.  But $^8$Be is unbound and will break up into
two $\alpha$ particles which are not $IMF$'s.  Thus $M_{IMF}$ drops to
zero.  In experiment this falls to about 0.3 rather than 0.  The simple
fact that $^6$Li (and an excited state of $^6$Li) is particle stable
whereas the states of $^8$Be are not is not embedded in our liquid-drop
model for ground state and Fermi-gas model for excited states.  Our
description gets better for larger nuclear systems but for very small
systems quantum mechanics of nuclear forces causes rapid changes in properties
as one goes from one excited state to another and one nucleus to another.
Our model can not accommodate this.

Let us go back to the case of $Z_s$=3 and treat it more realistically.
Using the abraison model,
when $Z_s$ is 3, PLF can have
$N_s$=3,4 and 5.  Probabilities for higher and lower values of $N_s$ are
small.  Following our model we get $M_{IMF}\approx 0.94$ with $Z_{bound}$
slightly less than 3.
When $Z_s$ is 4, significant probabilities
occur for$N_s$=5,6 and 7.  Following our model we get a small increase
in $M_{IMF}$ with $Z_{bound}$ whereas experimentally $M_{IMF}$ falls.
This discrepancy happens because in the Fermi-gas model there is
very little difference between properties of ground and excited states
of Li and Be whereas, in reality they are very different.  A much more
ambitious calculation for very small dissociating systems with $Z_s$ between
3 and 7 where we take binding energies and values of excited state energies
from experiments (this becomes more and more unwieldy as $Z_s$ increases)
is under way.

Fig.7 in ref \cite{Ogul} shows that SMM calculations are able to reproduce the
fluctuations faithfully.  Actually in those calculations the occurrences
of $Z_s,N_s$ with associated $E_x$ are not calculated but guessed so
that the ensemble produces the data as faithfully as possible.  For further
details how these calculations were done please refer to \cite{Botvina}.

\section{Towards a Universal Temperature Profile}
Knowing the temperature profile $T=T(b)$
in one case, say $^{124}$Sn on $^{119}$Sn, can we anticipate
what $T=T(b)$ will be like in another case, say, $^{58}$Ni on $^9$Be ?
In both the cases $b_{min}$ is zero and $b_{max}$ is $R_1+R_2$ yet we
can not expect the same functional form $T=T(b/b_{max})$ for both the cases.
In the first
case, near $b=0$ a small change in $b$ causes a large fractional change
in the mass of the PLF whereas, for $^{58}$Ni on $^9$Be,
near $b$=0, a small change in $b$ causes very little change in the mass
of the PLF.  Thus we might expect the temperature to change more rapidly in the
first case near $b$=0, whereas, in the second case, the temperature may
change very little since not much changed when $b$ changed a little.  In
fact, for Ni on Be, transport model calculations, HIPSE (Heavy Ion Phase Space
Exploration) and AMD (Antisymmetrised Molecular Dynamics) find that
starting from $b$=0, excitation energy/per particle changes very little
in the beginning \cite{Mocko2}.  In terms of our model, this would mean that for
Ni on Be, $T$ would be slow to change in the beginning.

We might argue that a measure of the wound that the projectile suffers in a
heavy ion collision is $1.0-A_s/A_0$ and that the temperature depends
upon the wound.  Thus we should expect $T=T(A_s(b)/A_0)$.  Just as we
can write $T(b)=C_0+C_1*b+C_2*b^2+...$ so also we could expand in powers
of $A_s(b)/A_0$,i.e., $T(b)=D_0+D_1(A_s(b)/A_0)+D_2(A_s(b)/A_0)^2+...$
We try such fits to the ``experimental'' temperature
profile given in Tables I and II.  From $b$ we deduce $A_s(b)/A_0$ and
plot $T$ as a function of $A_s(b)/A_0$.  A linear fit appears to be good
enough (Fig.5).

The specification that $T(b)=D_0+D_1(A_s(b)/A_0)$ has profound
consequences.  This means the temperature profile $T(b/b_{max})$ of
$^{124}$Sn on $^{119}$Sn is very different from that of $^{58}$Ni on $^9$Be.
In the first case $A_s(b)/A_0$ is nearly zero for $b=b_{min}$=0
whereas
in the latter case $A_s(b)/A_0$ is $\approx 0.6$ for $b=b_{min}$=0.
For $D_0$=7.5 MeV and $D_1$=-4.5 MeV, the temperature
profiles are compared in Fig.6.  Even more
remarkable feature is that the temperature profile of $^{58}$Ni on $^9$Be
is so different from the temperature profile of $^{58}$Ni on $^{181}$Ta.
In the latter case $b_{min}=R_{Ta}-R_{Ni}$ and beyond $b_{min}$,
$A_s(b)/A_0$ grows
from zero to 1 for $b_{max}$.
This is very similar to the temperature profile of $^{124}$Sn on $^{119}$Sn.

\section{Formulae for cross-sections}
Now that we have established that temperature $T$ should be considered
impact parameter $b$ dependent, let us write down how cross-sections
should be evaluated.  We first start with abrasion cross-section.  In eq.(1)
of \cite{Mallik},
the abrasion cross-section was written as
\begin{equation}
\sigma_{a,N_s,Z_s}=2\pi\int bdbP_{N_s,Z_s}(b)
\end{equation}
where $P_{N_s,Z_s}(b)$ is the probability that a PLF with $N_s$ neutrons and
$Z_s$ protons emerges in collision at impact parameter $b$.  Actually there
is an extra parameter that needs to be specified.  The complete labelling
is $\sigma_{a,N_s,Z_s,T}$ if we assume that irrespective of the value of $b$,
the PLF has a temperature $T$.  Here we have broadened this to the more general
case where the temperature is dependent on the impact parameter $b$.
Thus the PLF with $N_s$ neutrons and $Z_s$ protons will be formed in a
small range of temperature (as the production of a particular $N_s,Z_s$
occurs in a small range of $b$).

To proceed, let us discretize.  We divide the interval $b_{min}$ to
$b_{max}$ into small segments of length $\Delta b$.  Let the mid-point of
the $i$-th bin be $<b_i>$ and the temperature for collision at
$<b_i>$ be $T_i$.  Then
\begin{equation}
\sigma_{a,N_s,Z_s}=\sum_i\sigma_{a,N_s,Z_s,T_i}
\end{equation}
where
\begin{equation}
\sigma_{a,N_s,Z_s,T_i}=2\pi<b_i>\Delta bP_{N_S,Z_s}(<b_i>)
\end{equation}
PLF's with the same $N_s,Z_s$ but different $T_i$'s are treated independently.
The rest of the calculation proceeds as in \cite{Mallik}.  If, after abrasion, we
have, a system $N_s,Z_s$ at temperature $T_i$, CTM allows us to compute the
average population of the composite with neutron number $n$, proton number $z$
when this system breaks up (this composite is at temperature $T_i$).
Denote this by $M_{n,z}^{N_s,Z_s,T_i}$.  It then follows, summing over all
the abraded $N_s,Z_s$ that can yield $n,z$ the primary cross-section for
$n,z$ is
\begin{equation}
\sigma_{n,z}^{pr}=\sum_{N_s,Z_s,T_i}M_{n,z}^{N_s,Z_s,T_i}
\sigma_{a,N_s,Z_s,T_i}
\end{equation}
Finally, evaporation from these composites $n,z$ at temperatures $T_i$
is considered before comparing with experimental data.

\section{Cross-sections for different reactions}
We will now show some results for cross-sections using our model and compare
with experimental data.  We first show results for $^{124}$Sn on $^{119}$Sn
and $^{107}$Sn on $^{119}$Sn at 600 MeV/nucleon beam energy.  The experimental
data are plotted in \cite{Ogul} and the data were given to us, thanks to Prof.
Trautmann. The differential charge distributions and isotopic distributions for
$^{107}$Sn on $^{119}$Sn and $^{124}$Sn on $^{119}$Sn were theoretically calculated
using $T(b)=C_0+C_1b$ and and also $T(b)=D_0+D_1(A_s(b)/A_0)$.
So long as the temperature values at the two end points of $b$ are the same,
the answers did not differ much.  In Fig.7 we have shown results for
$T$ varying linearly with $b$ with $T_{max}=7.5$MeV and $T_{min}$=3 MeV.
At each $Z_{bound}$,
the charge distribution and isotopic distributions are calculated separately
and finally integrated over different $Z_{bound}$ ranges. The differential
charge distributions are shown in Fig.7 for different intervals of
$Z_{bound}/Z_0$ ranging between $0.0$ to $0.2$, $0.2$ to $0.4$, $0.4$ to $0.6$,
$0.6$ to $0.8$ and $0.8$ to $1.0$. For the sake of clarity the distributions
are normalized with different multiplicative factors. At peripheral collisions
(i.e. $0.8{\le}Z_{bound}/Z_0{\le}1.0$)  due to small temperature of the
projectile spectator, it breaks into one large fragment and small number of
light fragments, hence the charge distribution shows $U$ type nature. But with
the decrease of impact parameter the temperature increases, the projectile
spectator breaks into large number of fragments and the charge distributions
become steeper. In Figs.8 and 9 the integrated isotopic distributions
over the range $0.2{\le}Z_{bound}/Z_0{\le}0.8$ for Beryllium, Carbon, Oxygen
and Neon are  plotted and compared with the experimental result for
$^{107}$Sn on $^{119}$Sn and $^{124}$Sn on $^{119}$Sn reaction respectively.

Rest of the cross-sections shown all use $T(b)=7.5$MeV-$(A_S(b)/A_0)4.5$MeV.
First, in Fig. 10 the calculations of Fig. 7 are redone
but with the above parametrization.  Next we look at data for $^{58}$Ni on
$^9$Be and $^{181}$Ta at beam energy 140 MeV/nucleon done at Michigan State
University.  The data were made available to us by Dr.Mocko (Mocko,
Ph.D. thesis).  Calculations
were also done with $^{64}$Ni as beam.  Those results agree with experiment
equally well but are not shown here for brevity.
The results for $^{58}$Ni on $^9$Be and $^{58}$Ni on $^{181}$Ta are shown
in Figs.11 to 14.  The experimental data are from \cite{Mocko2}.
The chief difference from results shown in \cite{Mallik} is that we are able
 to include data for very peripheral collisions.  Next we look at some older
 data from $^{129}$Xe on $^{27}$Al at 790 MeV/nucleon \cite{Reinhold}.
Results are given in Figs. 15 and 16.

The parametrization $T(b)=7.5$MeV-$(A_s(b)/A_0)4.5$MeV was arrived at by
trying to fit many reaction cross-section data.  Better fit for $M_{IMF}$
vs. $Z_{bound}$ for Sn isotopes is found with slightly different
values: $T(b)=7.2$MeV-$(A_s(b)/A_0)3.2$MeV.

\section{Summary and Discussion}
We have shown that there are specific experimental data in projectile
fragmentation which clearly establish the need to introduce an impact
parameter dependence of temperature $T$ in the PLF formed.  Combining
data and a model one can establish approximate values of $T=T(b)$.
The model for cross-sections has been extended to incorporate this
temperature variation.  This has allowed us to investigate more peripheral
collisions.  In addition, the impact parameter dependence of temperature
appears to be very simple: $T(b)=D_0+D_1(A_s(b)/A_0)$
where $D_0$ and $D_1$ are constants, $A_s(b)$ is the mass
of the PLF and $A_0$ is the mass of the projectile.
With this model, we plan to embark upon an exhaustive study of available
data on projectile fragmentation.

\section{Acknowledgments}
This work was supported in part by Natural Sciences and Engineering Research
Council of Canada. The authors are thankful to Prof. Wolfgang Trautmann and Dr. M.
Mocko for access to experimental data. S. Mallik is thankful for a very productive and enjoyable
stay at McGill University for part of this work.  S. Das Gupta thanks Dr.
Santanu Pal for hospitality at Variable Energy Cyclotron Centre at Kolkata.

\begin{figure}
\includegraphics[height=3.75in,width=3.75in]{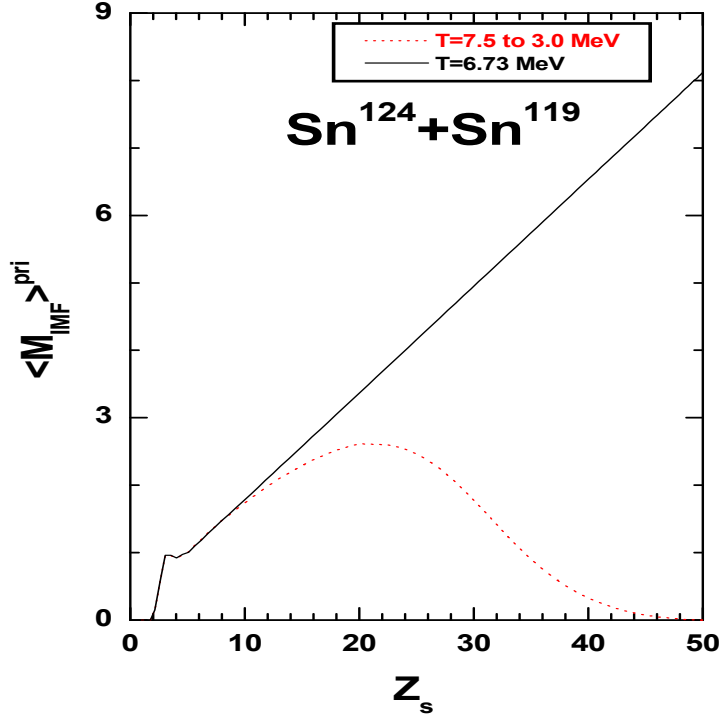}
\label{fig1}
\caption{ (Color Online) Mean multiplicity of intermediate-mass fragments $M_{IMF}$ (after multifragmentation stage), as a function of projectile spectator charge for $^{124}$Sn on $^{119}$Sn reaction calculated at a fixed temperature $T$=6.73 MeV (black solid line) and at a linearly decreasing temperature from 7.5 MeV at $b$=0 to 3 MeV at $b_{max}$ (red dotted line). The ordinate is labelled $<M_{IMF}>^{pri}$ as the effect of evaporation is not included. } \label{fig1}
\end{figure}

\begin{figure}
\includegraphics[height=5.5in,width=3.75in]{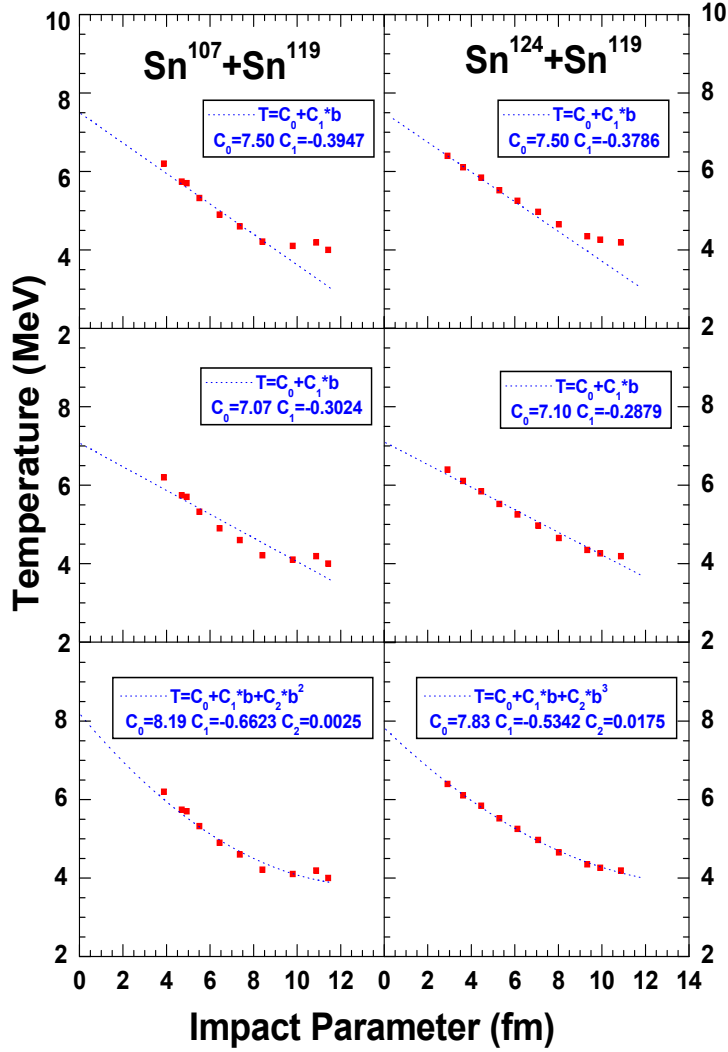}
\label{fig2}
\caption{ (Color Online) Impact parameter dependence of temperature for $^{107}$Sn on $^{119}$Sn (left panels) and $^{124}$Sn on $^{119}$Sn reactions (right panels). The red squires in the upper panels represent the extracted temperatures (sixth column of table-I and II) and the blue dotted lines are linearly decreasing temperature profile from 7.5 MeV to 3 MeV. The blue dotted lines of middle and lower panels represent fitting of extracted temperatures (red squires) with $T(b)=C_0+C_1*b$ and $T(b)=C_0+C_1*b+C_2*b^2$ equation respectively. The unit of $C_0$ is MeV, $C_1$ is MeVfm$^{-1}$ and $C_2$ is MeVfm$^{-2}$. }
\end{figure}

\begin{figure}
\includegraphics[height=3.25in,width=5.25in]{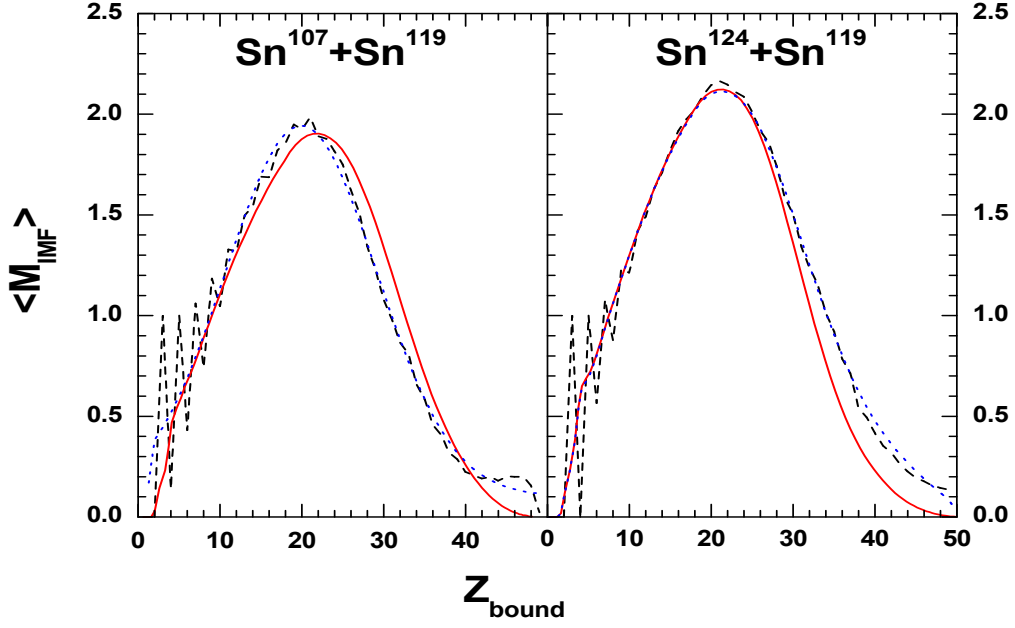}
\label{fig3}
\caption{ (Color Online) Mean multiplicity of intermediate-mass fragments $M_{IMF}$, as a function of $Z_{bound}$ for $^{107}$Sn on $^{119}$Sn (left panel) and $^{124}$Sn on $^{119}$Sn (right panel) reaction calculated using linearly decreasing temperature from 7.5 MeV to 3 MeV
(red solid lines) and $T(b)=C_0+C_1*b+C_2*b^2$ profile (blue dotted lines). The experimental results are shown by the black dashed lines. }
\end{figure}

\begin{figure}
\includegraphics[height=4.5in,width=6.0in]{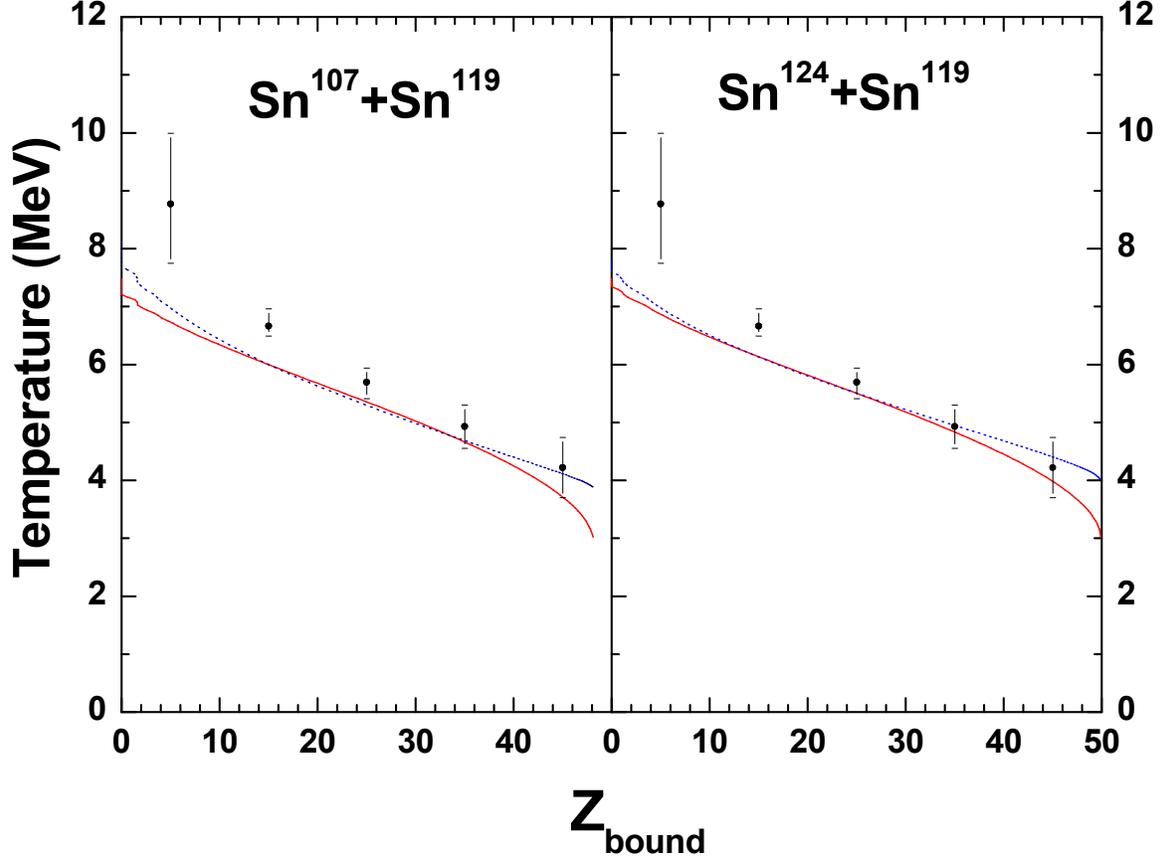}
\label{fig4}
\caption{ (Color Online) Comparison of theoretically used temperature profiles (i) temperature decreasing linearly with impact parameter from 7.5 MeV to 3 MeV (red solid lines), (ii)$T(b)=C_0+C_1*b+C_2*b^2$ fitting temperature (blue dotted lines) with that deduced by Albergo formula from experimental data (black points with error bars) for $^{107}$Sn on $^{119}$Sn (left panel) and $^{124}$Sn on $^{119}$Sn (right panel). }
\end{figure}

\begin{figure}
\includegraphics[height=5.25in,width=4.75in]{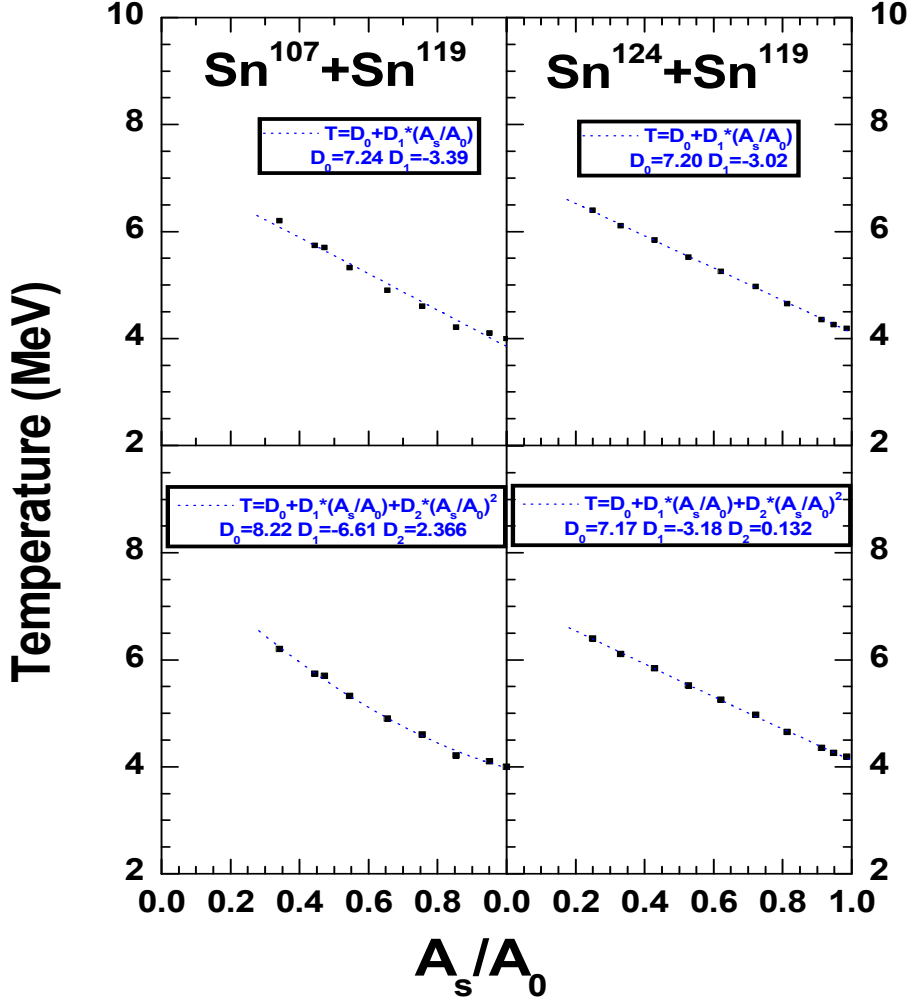}
\label{fig5}
\caption{ (Color Online) Fitting of extracted temperatures (red squires) with  $T(b)=D_0+D_1(A_s(b)/A_0)$ (blue dotted lines in upper panels) and $T(b)=D_0+D_1(A_s(b)/A_0)+D_2(A_s(b)/A_0)^2$ profile (blue dotted lines in lower panels) for $^{107}$Sn on $^{119}$Sn (left panels) and $^{124}$Sn on $^{119}$Sn (right panels). The units of $D_0$, $D_1$ and $D_2$ are MeV.}
\end{figure}

\begin{figure}
\includegraphics[height=4.0in,width=3.25in]{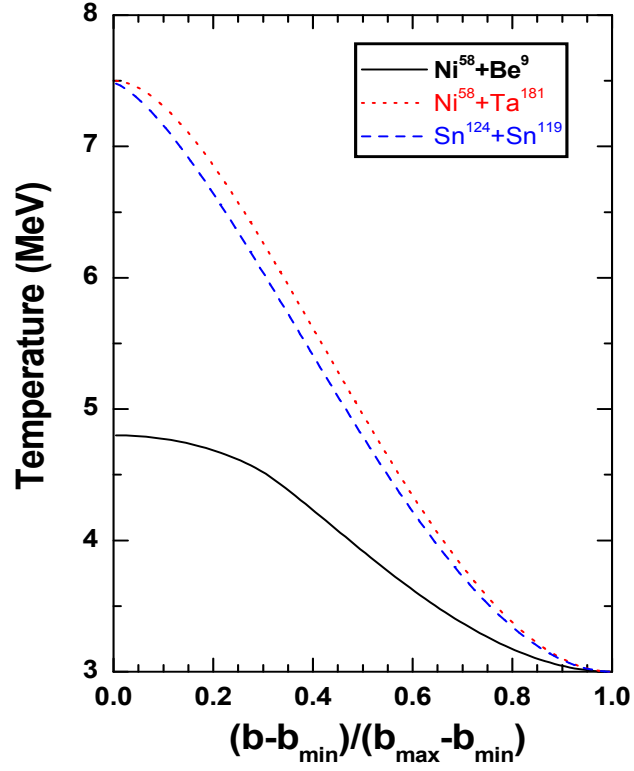}
\label{fig6}
\caption{ (Color Online) Temperature profile for $^{58}$Ni on $^9$Be (black solid line), $^{58}$Ni on $^{181}$Ta (red dotted line) and $^{124}$Sn on $^{119}$Sn (blue dashed line) by considering $T=7.5-4.5(A_s(b)/A_0)$}
\end{figure}

\begin{figure}
\includegraphics[height=4.5in,width=6.00in]{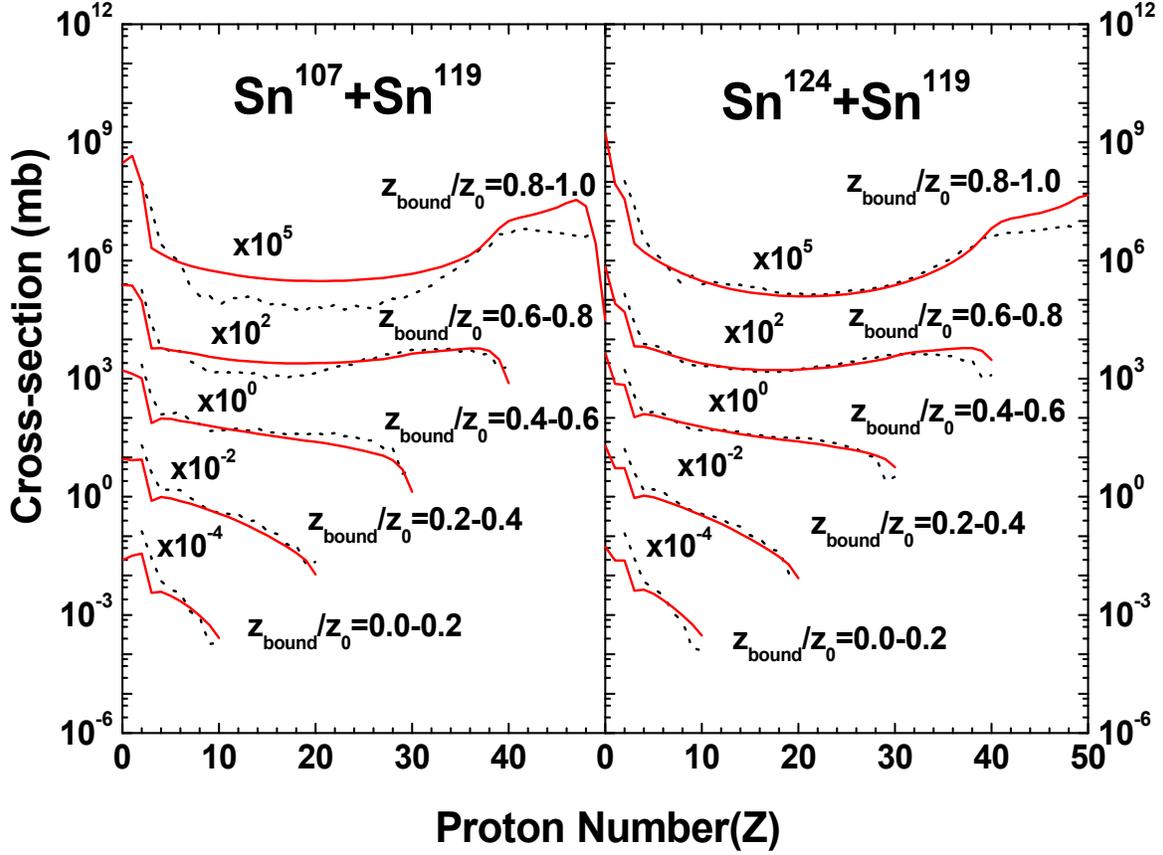}
\label{fig7}
\caption{ (Color Online) Theoretical total charge cross-section
distribution (red solid lines) for $^{107}$Sn on
$^{119}$Sn (left panel) and $^{124}$Sn on
$^{119}$Sn reaction (right panel) sorted into five intervals of $Z_{bound}/Z_0$ ranging between $0.0$ to $0.2$, $0.2$ to $0.4$, $0.4$ to $0.6$, $0.6$ to $0.8$ and $0.8$ to $1.0$ with different multiplicative factors $10^{-4}$, $10^{-2}$, $10^{0}$, $10^{2}$, $10^{5}$ respectively. The experimental data are shown by black dashed lines. Theoretical calculation is done using linearly decreasing temperature from 7.5 MeV at $b$=0 to 3 MeV at $b_{max}$.}
\end{figure}

\begin{figure}
\includegraphics[height=4.5in,width=5.00in]{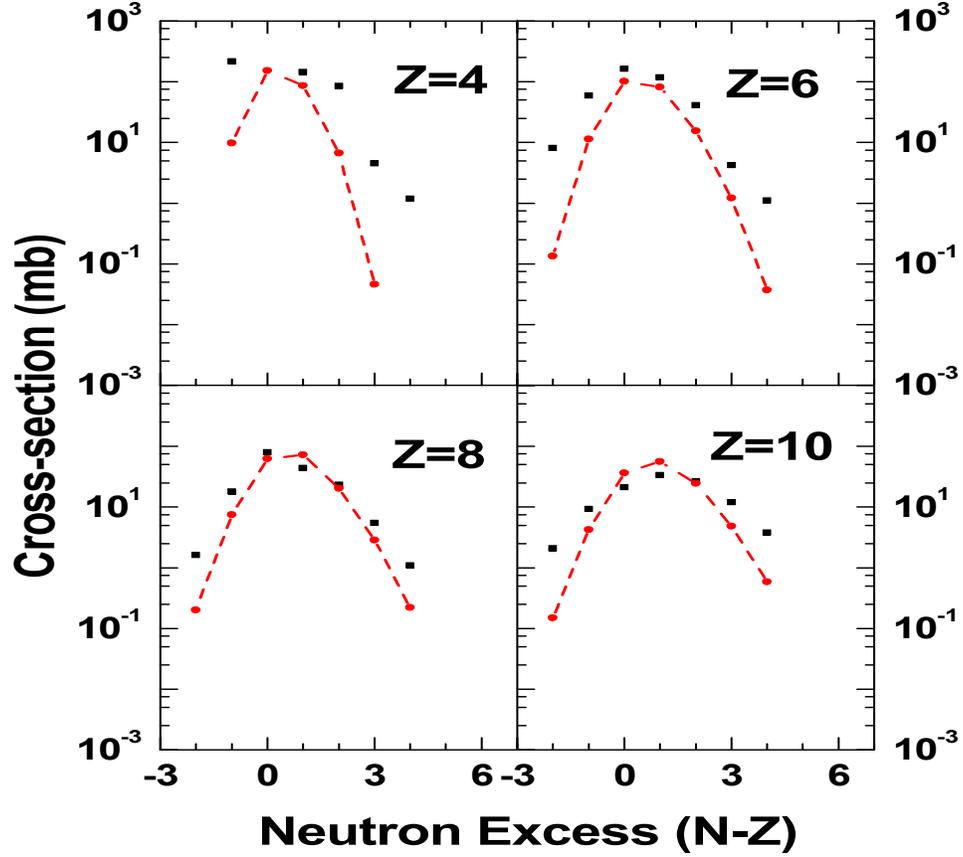}
\label{fig8}
\caption{ (Color Online) Theoretical isotopic cross-section
distribution (circles joined by dashed lines) for $^{107}$Sn on
$^{119}$Sn reaction summed over $0.2{\le}Z_{bound}/Z_0{\le}0.8$. The experimental data are shown by black squires.  Theoretical calculation is done using linearly decreasing temperature from 7.5 MeV at $b$=0 to 3 MeV at $b_{max}$.}
\end{figure}

\begin{figure}
\includegraphics[height=4.5in,width=5.00in]{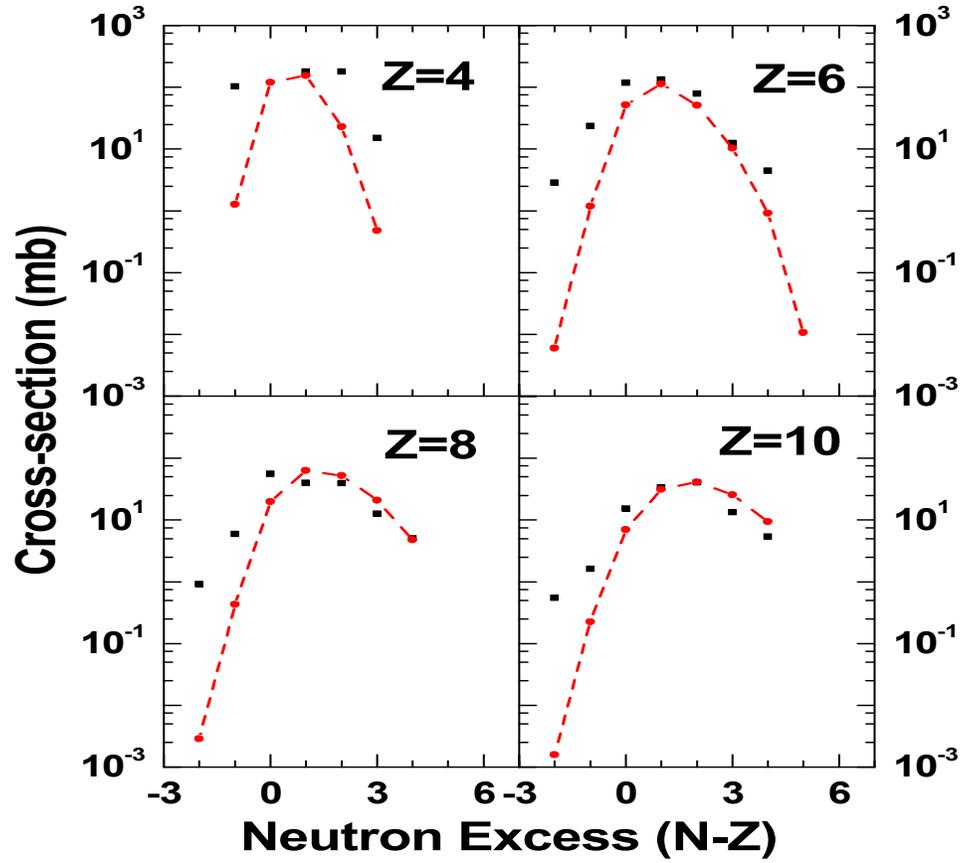}
\label{fig9}
\caption{ (Color Online) Same as Fig. 8, except that here the projectile is $^{124}$Sn instead of $^{107}$Sn.}
\end{figure}

\begin{figure}
\includegraphics[height=4.5in,width=6.00in]{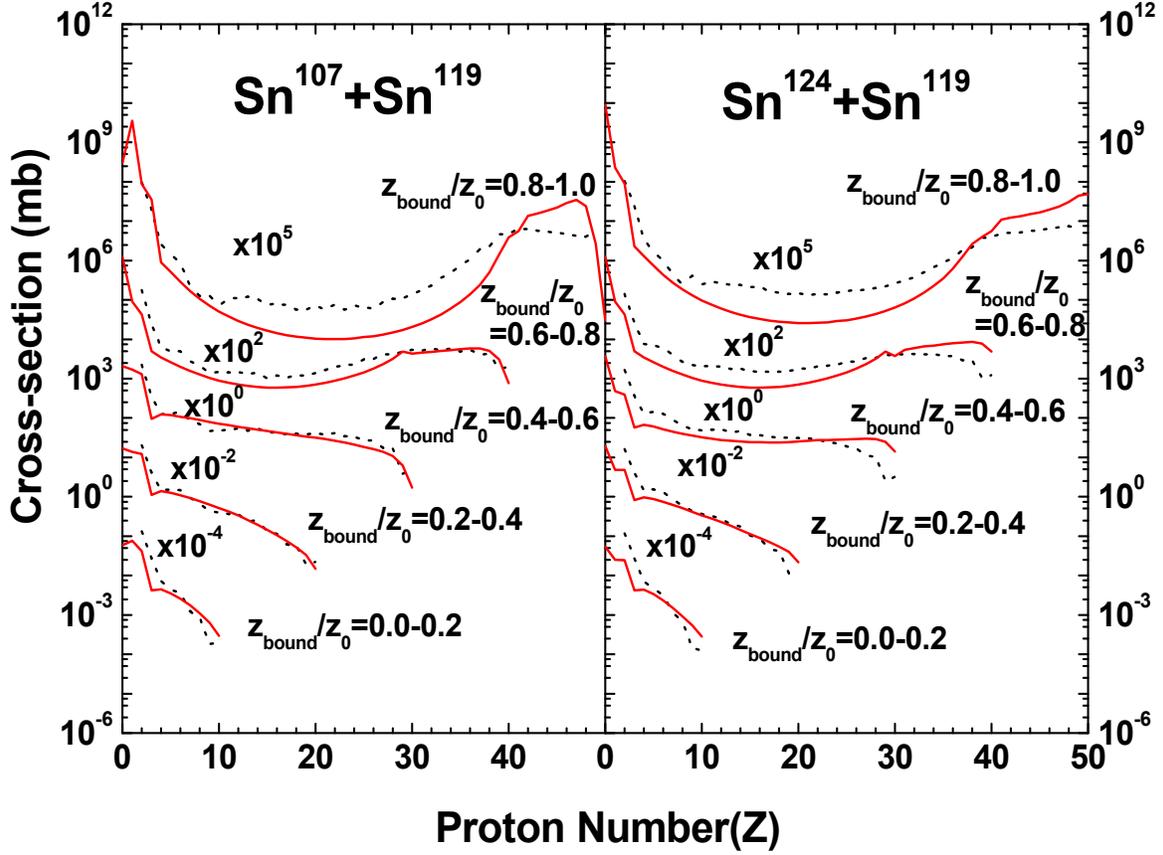}
\label{fig10}
\caption{ (Color Online) Same as Fig. 7 except that here the temperature profile is $T(b)=7.5MeV-(A_S(b)/A_0)4.5MeV$ instead of linearly decreasing temperature from 7.5 MeV at $b$=0 to 3 MeV at $b_{max}$ }
\end{figure}

\begin{figure}
\includegraphics[width=6.0in,height=4.5in,clip]{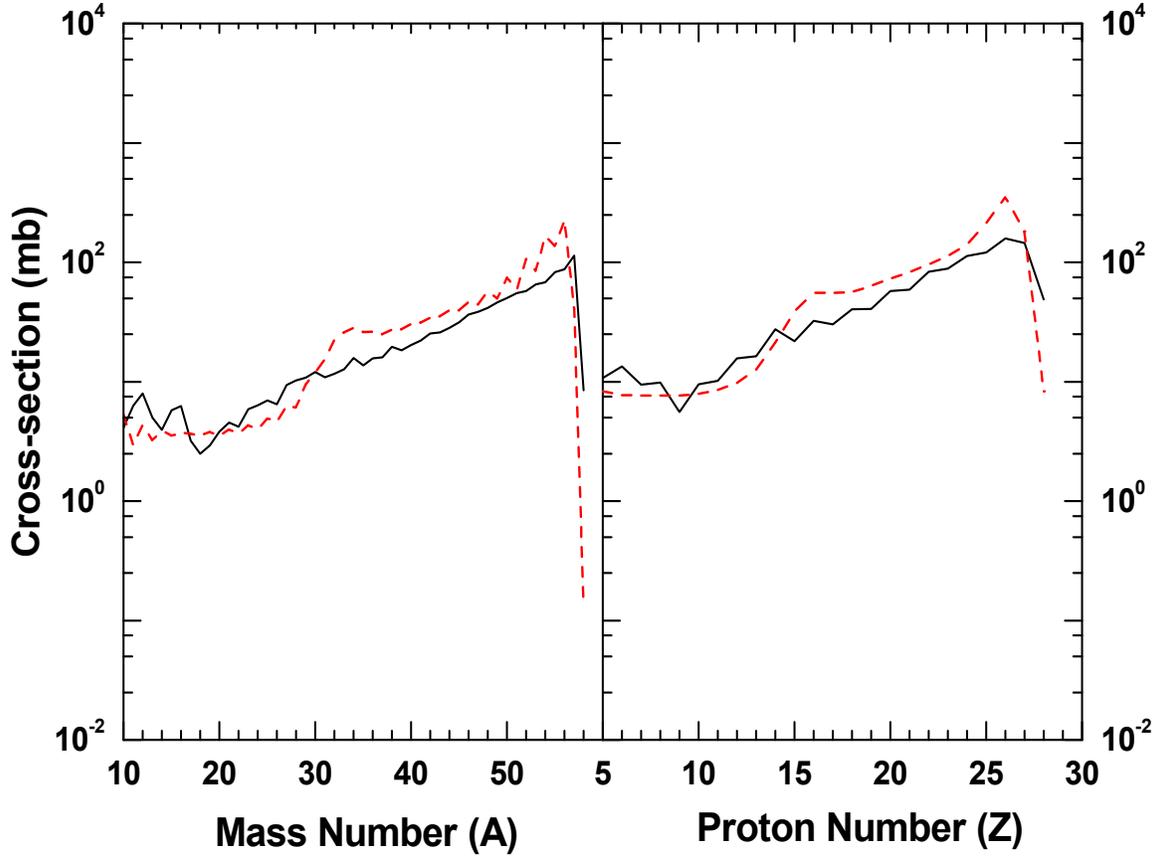}
\caption{ (Color Online) Total mass (left panel) and total charge
(right panel) cross-section distribution for the $^{58}$Ni on
$^{9}$Be reaction. The left panel shows the cross-sections as a
function of the mass number, while the right panel displays the
cross-sections as a function of the proton number. The theoretical
calculation is done using temperature decreasing linearly with $A_s/A_0$ from 7.5 MeV to 3.0 MeV (dashed line) and compared with the experimental data (solid line). } \label{fig11}
\end{figure}

\begin{figure}
\includegraphics[width=6.0in,height=4.5in,clip]{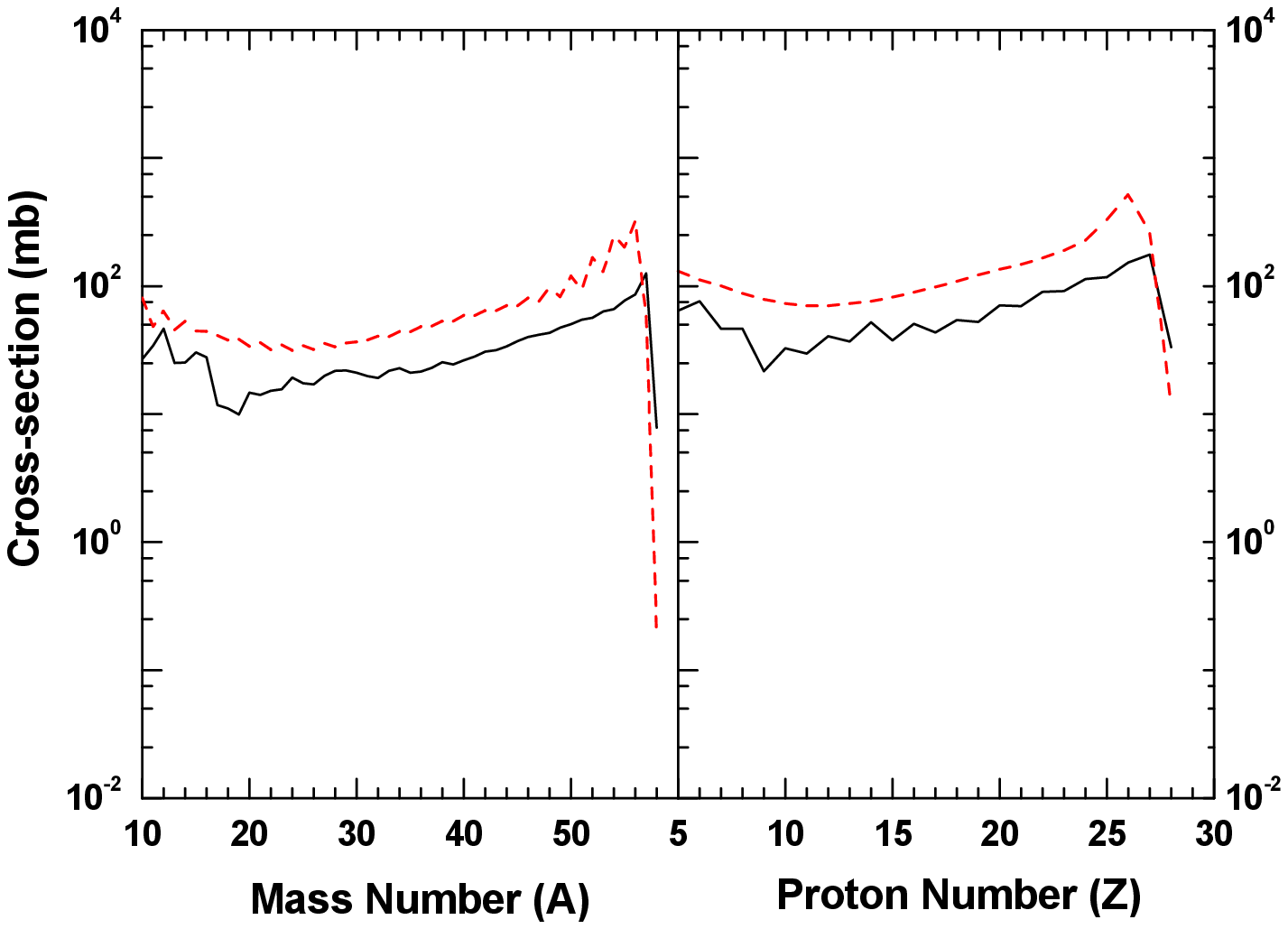}
\caption{ (Color Online)  (Color Online) Same as Fig. 11 except that here the target
is $^{181}$Ta  instead of $^{9}$Be.} \label{fig12}
\end{figure}

\begin{figure}
\includegraphics[width=6.0in,height=4.5in,clip]{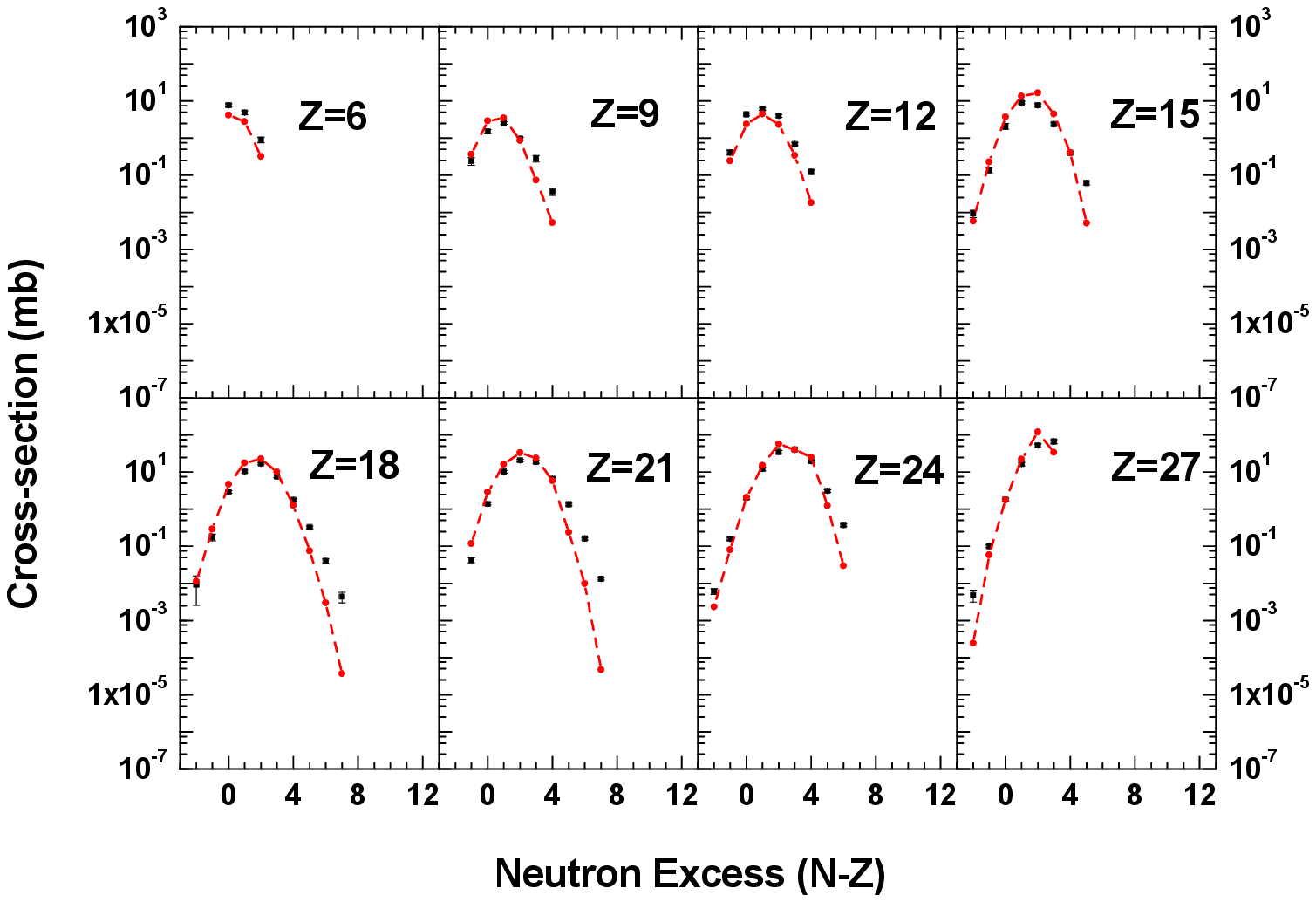}
\caption{ (Color Online) Theoretical isotopic cross-section
distribution (circles joined by dashed lines) for $^{58}$Ni on
$^{9}$Be reaction compared with experimental data (squares with
error bars).} \label{fig13}
\end{figure}

\begin{figure}
\includegraphics[width=6.0in,height=4.5in,clip]{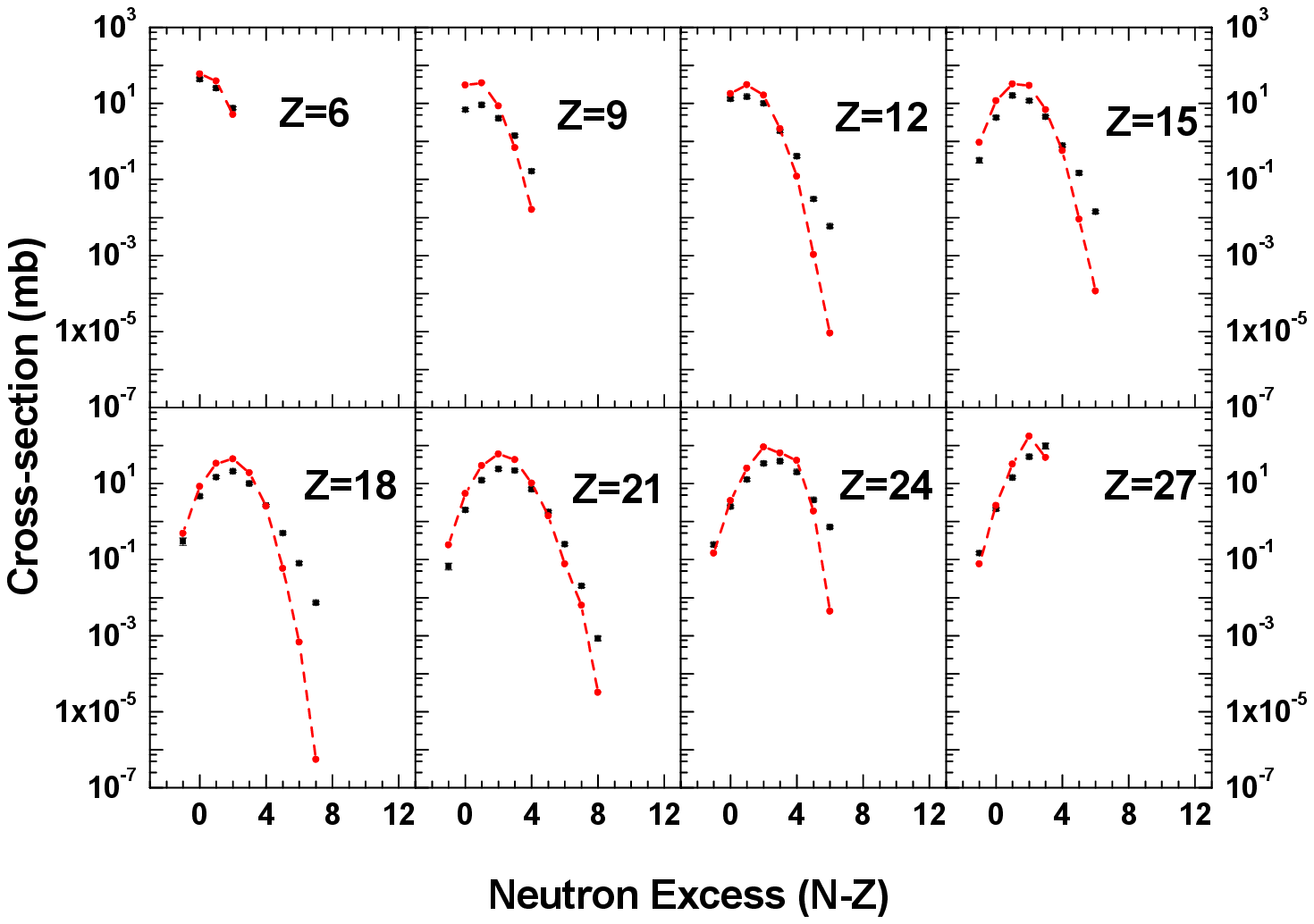}
\caption{ (Color Online) Same as Fig. 13 except that here the target
is $^{181}$Ta  instead of $^{9}$Be.} \label{fig14}
\end{figure}

\begin{figure}
\includegraphics[width=3.0in,height=4.0in,clip]{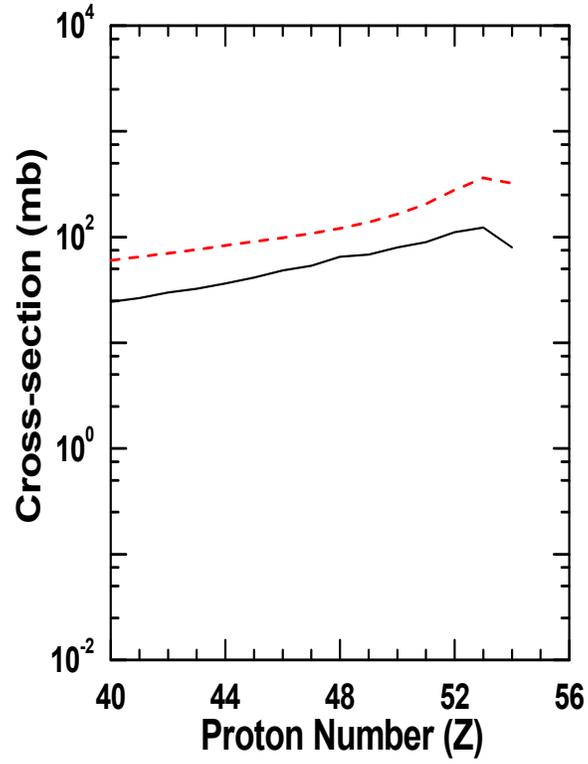}
\caption{ (Color Online) Total charge cross-section distribution for
the $^{129}$Xe on $^{27}$Al reaction. The theoretical
calculation is done using temperature decreasing linearly with $A_s/A_0$ from 7.5 MeV to 3.0 MeV (dashed line) and compared with the experimental data (solid line).}
\label{fig15}
\end{figure}

\begin{figure}
\includegraphics[width=6.0in,height=4.5in,clip]{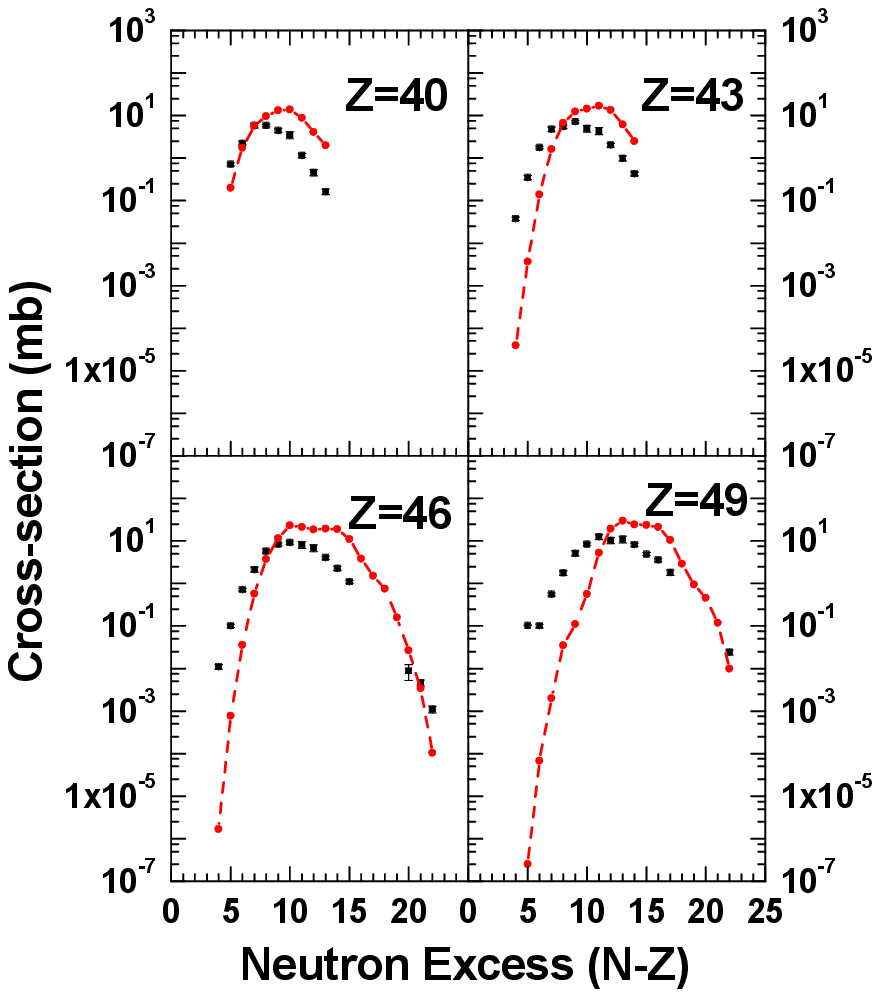}
\caption{ (Color Online) Theoretical isotopic cross-section
distribution (circles joined by dashed lines) for $^{129}$Xe on
$^{27}$Al reaction compared with experimental data (squares with
error bars).} \label{fig16}
\end{figure}

\end{document}